\documentclass{PHYEAUTH}
\usepackage{graphicx}
\usepackage{amsmath}
\usepackage{amssymb}

\newcommand{\beq}   {\begin{equation}}
\newcommand{\eeq}   {\end{equation}}
\newcommand{\ba}   {\begin{eqnarray}}
\newcommand{\ea}   {\end{eqnarray}}

\newcommand{\dotdot}   {\mbox{\tiny Dot-Dot}}
\newcommand{\dotleads}   {\mbox{\tiny Dot-Leads}}
\newcommand{\lead}   {\mbox{\tiny Leads}}

\begin{document}

\begin{frontmatter}

\title{Transmission in double quantum dots in the Kondo regime:
Quantum-critical transitions and interference effects}

\author[address1]{Luis G.~G.~V. Dias da Silva \thanksref{thank1}},
\author[address1]{Nancy Sandler}
\author[address2]{Kevin Ingersent}
and
\author[address1]{Sergio E. Ulloa}

\address[address1]{Department of Physics and Astronomy, Nanoscale and
Quantum Phenomena Institute, Ohio University, Athens, Ohio
45701--2979}

\address[address2]{Department of Physics, University of Florida, P.O.\
Box 118440, Gainesville, Florida, 32611--8440}

\thanks[thank1]{
Corresponding author. E-mail: dias@phy.ohiou.edu}

\begin{abstract}

We study the transmission through a double quantum-dot system in
the Kondo regime. An exact expression for the transmission
coefficient in terms of fully interacting many-body Green's
functions is obtained. By mapping the system into an effective
Anderson impurity model, one can determine the transmission using
numerical renormalization-group methods. The transmission exhibits
signatures of the different Kondo regimes of the effective model,
including an unusual Kondo phase with split peaks in the spectral
function, as well as a pseudogapped regime exhibiting a quantum
critical transition between Kondo and unscreened phases.

\end{abstract}

\end{frontmatter}

\section{Introduction}

Semiconductor quantum dots have played a major role in the
investigation of strongly correlated effects in nanoscale systems,
as highlighted in the pioneering experiments on Kondo effect in
these devices \cite{Goldhaber-Gordon:156:1998}. More recently,
\textit{double-dot} setups have been used to investigate
two-impurity and two-channel Kondo physics
\cite{DQDExpts}. These experimental developments
have spurred great theoretical interest in double dots in the
Kondo regime. In particular, studies of side-coupled
\cite{DQDSideDot_Theory}
and parallel
\cite{DQDParallel_Theory} dot configurations have been carried
out.

Motivated by these accomplishments, we study transport properties
of double quantum dots (DQDs) with one interacting dot  (``dot 1")
coupled to a large dot, effectively noninteracting (``dot 2", with
energy $\varepsilon_2$). This seemingly simple setup can be
described by an Anderson model coupled to a fermionic bath with a
nonconstant density of states (DoS). Compared with the standard
case of a metallic host (constant DoS), a nonconstant DoS leads to
nontrivial new and interesting features in the many-body Kondo
state.

As shown previously \cite{Silva:096603:2006}, a {\em splitting}
the Kondo resonance appears when the DoS shows a sharp peak at the
Fermi energy, while the Kondo singlet itself is preserved. The
proposed DQD setup also allows for the appearance of a pseudogap
in the effective DoS, leading to a critical transition between
Kondo and non-Kondo phases.  These phenomena substantially modify
the spectral function of the interacting dot. We show here that
another quantity, the transmission coefficient, can also be used
to explore the critical transition and other features of this
system.

We obtain an exact expression for the transmission coefficient
$T(\omega$) in terms of the fully interacting many-body Green's
functions for the DQD system, which are then calculated within the
numerical renormalization-group (NRG) framework. This approach
allows us to fully describe the characteristics of the
transmission coefficient in the different regimes mentioned above.

\section{Model }

\begin{figure}[h]
\centerline{\includegraphics*[height=0.4\columnwidth,width=0.6\columnwidth]{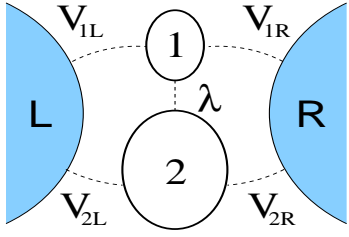}}
\caption{\label{fig:DDot}
Schematic representation of the double-dot system. }
\end{figure}

The model describes the DQD setup depicted in Fig. \ref{fig:DDot}:
quantum dots $1$ (singly occupied with charging energy $U$ and
energy $\varepsilon_1$) and $2$ (energy $\varepsilon_2$) are
coupled to metallic leads ($\ell=L,R$) and to each other.

The Hamiltonian is given by $H = H_{1} + H_{2} + H_{\lead} +
H_{\dotdot} + H_{\dotleads}$, where
\begin{eqnarray}
H_{\lead} & = & \sum_{{\bf k},\,\ell=L,R} \varepsilon_{\ell k}
c^{\dagger}_{\ell {\bf k} \sigma}
c^{\phantom{\dagger}}_{\ell {\bf k} \sigma}
\, ,\nonumber\\
H_{i=1,2} & = & \varepsilon_i c^{\dagger}_{i \sigma}
c^{\phantom{\dagger}}_{i \sigma} +
\delta_{i,1} U n_{1 \uparrow}n_{1 \downarrow} \, ,\nonumber\\
H_{\dotleads} & = & \sum_{{\bf k},\,\ell=L,R}
\left(
V_{i \ell {\bf
k}}c^{\dagger}_{i \sigma} c^{\phantom{\dagger}}_{\ell {\bf k}
\sigma}+V^{*}_{i \ell {\bf k}\sigma}
c^{\dagger}_{\ell {\bf k} \sigma} c^{\phantom{\dagger}}_{i \sigma}
\right)
,\nonumber\\
H_{\dotdot} & = &
\lambda c^{\dagger}_{1 \sigma} c^{\phantom{\dagger}}_{2 \sigma} +
\lambda^* c^{\dagger}_{2 \sigma} c^{\phantom{\dagger}}_{1 \sigma}
\, .
\label{Eq:Hamiltonian}
\ea

The transmission amplitude for an electron tunneling from the left
lead to the right lead can be obtained from the retarded Green's
function connecting the leads, $G_{L{\bf k}R{\bf k}^\prime}(\omega)
\equiv \langle\langle c^{\phantom{\dagger}}_{L{\bf k}}, \,
c^{\dagger}_{R{\bf k}^\prime} \rangle \rangle$:
\beq
G_{L{\bf k}R{\bf k}^\prime}(\omega) = \sum_{i,j=1,2} g_{L{\bf k}}
\; V^*_{iL {\bf k}}  G_{ij}(\omega) V_{jR {\bf k}^\prime} \;
g_{R{\bf k}^\prime} \, .
\eeq
with $g_{\ell{\bf k}}\equiv (\omega^+-\varepsilon_{\ell k})^{-1}$.
In the derivation of the expression above, we have used identities
from the equations of motion for fermionic operators
$c^{\phantom{\dagger}}_a, \, c^{\dagger}_b$ in the frequency
domain that include \cite{Caio}:
\ba
\omega \, \langle\langle c^{\phantom{\dagger}}_a:c^{\dagger}_b
\rangle\rangle_{\omega} &=& \langle [ c^{\phantom{\dagger}}_a,
c^{\dagger}_b ]_\eta \rangle + \langle\langle [
c^{\phantom{\dagger}}_a, H]_{-}:c^{\dagger}_b
\rangle\rangle \; ,\nonumber\\
\omega \, \langle\langle c^{\phantom{\dagger}}_a:c^{\dagger}_b
\rangle\rangle_{\omega} &=& \langle [ c^{\phantom{\dagger}}_a,
c^{\dagger}_b ]_\eta \rangle - \langle\langle c^{\phantom{\dagger}}_a:
[ c^{\dagger}_b,H ]_{-} \rangle\rangle \, .
\label{Eq:EOMs}
\ea

In the wide-band limit $D \gg |\omega|$, where $D$ is the
half-bandwidth, we approximate $V_{i \ell {\bf k}}\equiv V_{i
\ell}$ (${\bf k}$-independent couplings) and $\sum_{{\bf
k}}(\omega^+ - \varepsilon_{k})^{-1} \rightarrow -i\pi \rho_0$
(where the same DoS $\rho_0$ is assumed for each lead) and one
obtains the following expression for the energy-dependent
transmission from $L$ to $R$:
\begin{eqnarray}
T(\omega)  &=&  \sum_{{\bf k},{\bf k}^\prime} G_{L{\bf k}R{\bf
k}^\prime}(\omega) =  2 \pi \rho_0 \! \! \! \! \sum_{i,j=1,2} \!
\! V^*_{iL} G_{ij}(\omega) V_{jR} \, .
\end{eqnarray}

In the following, we assume a symmetric configuration
$V_{iR}=V_{iL}=V_i/\sqrt{2}$ with real values for the couplings
for simplicity. Defining $\Delta_i \equiv \pi \rho_0 V^2_i$ and
$\Delta_{12} \equiv \pi \rho_0 V_1 V_2$, $T(\omega)$ can be
written in a compact form:
\begin{eqnarray}
T(\omega)  &=&  \Delta_1 G_{11} + \Delta_2 G_{22} + \Delta_{12}
\left( G_{12} + G_{21} \right) .
\label{Eq:TexpII}
\end{eqnarray}

All $G_{ij}$ in (\ref{Eq:TexpII}) are fully interacting Green's
functions. Interaction effects are introduced into the dot-2
Green's function by direct and indirect (via the leads) tunneling
to dot 1. The identities (\ref{Eq:EOMs}) can be used to establish
the relations
\begin{eqnarray}
G_{22}(\omega) & = & G^{(0)}_{22}(\omega) + G_{21}(\omega)
G_{11}(\omega)G_{12}(\omega) \, ,  \nonumber\\
G_{12}(\omega) & = & G^{(0)}_{22}(\omega) \left( \lambda -
i\Delta_{12} \right) G_{11}(\omega) \, .
\end{eqnarray}
Notice that for real couplings, $G_{12}(\omega)=G_{21}(\omega)$.
These relationships in turn lead to an \textit{exact} expression
for the transmission involving only the noninteracting dot-2
Green's function $G^{(0)}_{22}(\omega)=\left(\omega-\varepsilon_2 + i
\Delta_2 \right)^{-1}$ and the fully interacting dot-1 Green's
function $G_{11}(\omega)\equiv \langle\langle
c^{\phantom{\dagger}}_{1 \sigma}\!:\!
c^{\dagger}_{1 \sigma} \rangle \rangle$:
\begin{eqnarray}
T(\omega) & = &\Delta_1 G_{11}(\omega) + 2 \Delta_{12}
\left[ G^0_{22}(\omega) (\lambda - i\Delta_{12})G_{11}(\omega) \right]
\nonumber \\
+ & \Delta_2 &   G^0_{22}(\omega) \left[ 1 + G^0_{22}(\omega) (\lambda -
i\Delta_{12})^2 G_{11}(\omega) \right] .
\label{Eq:Tfinal}
\end{eqnarray}

The dot-1 Green's function $G_{11}$ is obtained from NRG
calculations in the following manner: Hamiltonian
(\ref{Eq:Hamiltonian}) is mapped
\cite{Silva:096603:2006}
onto an Anderson impurity
connected to an effective nonconstant
DoS through a hybridization function
\begin{equation}
\label{Eq:Delta}
  \Delta(\omega) = \pi \rho_2(\omega) \left[\lambda + (\omega -
    \varepsilon_2)
    \sqrt{\Delta_1/\Delta_2} \,
    \right]^2,
\end{equation}
with $\rho_2(\omega)=\Delta_2/\{\pi[(\omega-\varepsilon_2)^2 +
\Delta_2^2]\}$. In general, $\Delta(0)\neq0$ gives the effective
broadening of the single-particle Hubbard peaks in dot 1 produced
by the coupling to the effective nonconstant DoS. Notice that, for
$\Delta_2,\lambda \rightarrow 0$ (single-dot case), $\Delta(0)
\rightarrow \Delta_1$.

We focus on two limiting configurations. In the ``side-dot'' limit
($\Delta_1=0; \Delta_2, \lambda \neq 0$), dot 1 is connected to the
leads only by second-order tunneling processes mediated by dot 2,
and $\Delta(\omega)$ has a Lorentzian form, with a peak of
width $\Delta_2$ centered at $\varepsilon_2$. In the ``parallel-dot''
limit ($\Delta_1, \Delta_2 \neq 0; \lambda = 0$), dot 1 and
dot 2 are connected only by indirect tunneling through the leads.
Now $\Delta(\omega)$ has an ``inverted Lorentzian" shape, vanishing
as $|\omega-\varepsilon_2|^2$ at $\omega=\varepsilon_2$; if dot 2 is
in resonance with the leads ($\varepsilon_2=0$, the case assumed
throughout the remainder of the paper), this corresponds to a
pseudogap in the effective DoS with exponent $2$.

 The spectral function $A_{11}(\omega) = - \pi^{-1} \mbox{Im }
G_{11}(\omega)$ is obtained from the NRG spectra using the method
described in Ref. \cite{Bulla:045103:2001}. In short, at each NRG
iteration $N$, one collects partial information for the spectral
function by approximating $A_{11}(\omega) = A^N_{11}(\omega)$ for
$\omega_N < \omega \lesssim 10\omega_N$ where $\omega_N \approx D
\Lambda^{-N/2}$ ($\Lambda >1$) is the characteristic energy scale
at iteration $N$. A continuous function can be obtained by
replacing the $\delta$-functions at energies $E_n$ entering the
Lehmann expression by logarithmically broadened lines
$\delta^b(\omega - E_n) \propto
\exp\left[-(\ln{\omega}-\ln{E_n})^2/b^2\right]$.

\begin{figure}[t]
\centerline{\includegraphics*[height=0.7\columnwidth,width=0.9\columnwidth]{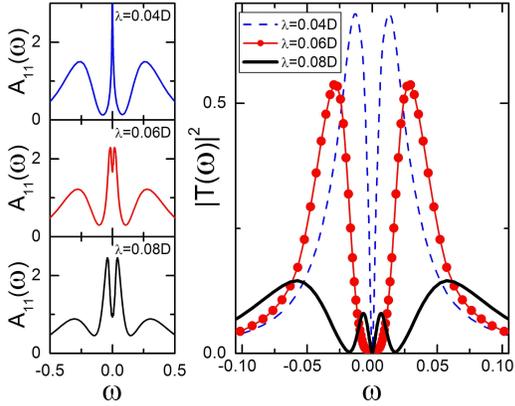}}
\caption{ (color online) Spectral densities (left panels) and
transmission amplitude (right panel) for the side dot
configuration with $\varepsilon_1=-U/2$ and $\lambda=0.04D$ (dashed
lines), $0.06D$ ($\bullet$) and $0.08D$ (thick solid line). Notice
the appearance of secondary peaks in $|T(\omega)|^2$ as $\lambda$
increases.}
\label{fig:TwSideDot}
\end{figure}
Next, $\mbox{Re } G_{11}(\omega)$ is obtained from
$A_{11}(\omega)$ via the appropriate Kramers-Kronig (KK)
transformation. Because of the subtleties involved in the use of
such transformations, we compared (i) the results of transforming
$A_{11}(\omega)$ obtained numerically as described above with (ii)
those from performing the KK at each NRG step, calculating
$F_{N}(\omega)= \pi^{-1} \mathcal{P} \int d\omega^{\prime}
A^N_{11}(\omega^\prime)/\left(\omega^\prime - \omega\right)$ and
then approximating $\mbox{Re } G_{11}(\omega) = -\pi^{-1}
F_N(\omega)$ for $\omega_N < \omega \lesssim 10\omega_N$. We
obtained qualitatively similar results with both methods, with
better agreement for larger $b$ entering $\delta^b(\omega - E_n)$
($b=0.3$--$0.6$ for the results presented). For large $N$, an
additional, broadening-independent estimate of $G_{11}(0)$ can be
obtained from the expression \cite{Zhu} $F_N(0) = Z_N(0)^{-1}
\sum_n \left( |\langle n | c_{1 \sigma} | 0 \rangle|^2 - |\langle
0 | c_{1 \sigma} | n \rangle|^2 \right) / E^N_n$, where $E^N_n$,
$|n \rangle_N$ and $Z_N(0)$ are, respectively, the eigenenergies,
eigenstates and zero-temperature grand-canonical partition
function at NRG iteration $N$, with $n=0$ representing the ground
state.

\section{Results}

In the following, we take $U=0.5D$ and $\Delta_2=0.02D$. In the
side-dot configuration considered, the effective dot-1
hybridization $\Delta(\omega)$ has a resonance of width $\Delta_2$
at the Fermi energy. Note that $\Delta(0)=\lambda^2/\Delta_2$, so
that the Kondo temperature $T_K \propto e^{-1/\Delta(0)}$ will
increase as $\lambda$ increases.

As previously shown \cite{Silva:096603:2006}, the presence of a
resonance in the effective hybridization leads to a splitting in
the spectral function for $\lambda \gtrsim \lambda^*$, where
$\lambda^*$ is defined implicitly by $T_K(\lambda^*,\Delta_2) =
\Delta_2$. This can be seen in the left panels of Fig.\
\ref{fig:TwSideDot}, for which $\lambda^* \sim 0.05D$.

The transmission amplitude is shown in the main panel of Fig.\
\ref{fig:TwSideDot}. At $\omega=0$, the Friedel sum rule requires
that $\mbox{Im} G_{11}(0)=-\Delta_2/\lambda^2$. Substituting this
result in Eq.\ (\ref{Eq:Tfinal}) leads to $|T(0)|^2=0$
\textit{independently of $\lambda,\Delta_2$}. The vanishing of
transmission across the double-dot system is essentially an
interference effect: the path going directly from the left lead to
the right lead through dot 2 interferes destructively with the
alternate path in which the electron tunnels in and out of dot 1.

Signatures of the splitting in the spectral density appear in the
transmission amplitude at finite $\omega$. For increasing values
of $\lambda$, the onset of the splitting in the spectral function
(left panels in Fig.\ \ref{fig:TwSideDot}) is accompanied by the
appearance of satellite peaks in $|T(\omega)|^2$ near $\omega=0$.

\begin{figure}[t]
\centerline{\includegraphics*[height=0.7\columnwidth,width=0.9\columnwidth]{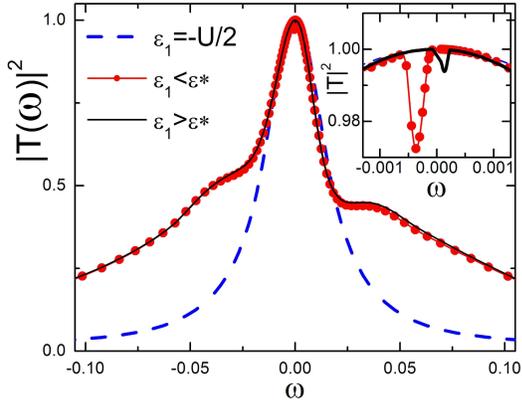}}
\caption{ (color online) Transmission amplitude for the parallel
configuration with $\Delta_1=0.05D$, and $\varepsilon_1=-U/2$ (dashed
line) $-0.088U$ ($\bullet$) and $-0.086U$. The critical point is
reached at $\varepsilon^*_1=-0.086524U$. Inset: A dip in $T(\omega)$
crosses $\omega=0$ due to the pseudogap-induced transition. }
\label{fig:TwParallel}
\end{figure}

In the case of the dots in a parallel configuration, the
situation is different, as shown in Fig.\ \ref{fig:TwParallel}.
Notice that $|T(0)|^2=1$ always, which is a direct consequence of
the presence of the pseudogap in the hybridization
function $\Delta(\omega)$. In the resulting effective pseudogapped Anderson model,
the spectral density vanishes as $|\omega|^2$ at the Fermi energy.
Thus, $\mbox{Im }G_{11}(0)=0$, effectively closing the
transmission through dot 1 at the Fermi energy. In this regime,
the transmission is dominated by the resonant tunneling through dot 2.

Important differences appear in particle-hole (p-h) symmetric and
asymmetric regimes. In the p-h symmetric case ($\varepsilon_1=-U/2$,
dashed lines in Fig.\ \ref{fig:TwParallel}) the dot-1 spectral
density varies as $\omega^2$ within a relatively large range
$|\omega| < \Delta_2$ around the Fermi energy
\cite{Silva:096603:2006}. In this case, $|T(\omega)|^2$ is
essentially a Lorentzian of width $\Delta_2$.

Away from p-h symmetry, a quantum critical point separating Kondo
and non-Kondo phases can be reached \cite{Silva:096603:2006}.
Passage through the quantum critical point is reflected in
position of a peak in the dot-1 spectral function at a frequency
$\omega^*$ that crosses from $\omega^*>0$ in the Kondo phase to
$\omega^* < 0$ in the unscreened phase \cite{Vojta:014511:2001}.

We find that the transition also has a signature in the
transmission. The inset to Fig.\ \ref{fig:TwParallel} shows that
the low-energy peak in $A_{11}$ translates into a \textit{dip} in
$|T(\omega)|^2$ at $\omega\approx\omega^*$---a dip that passes
through the Fermi energy at the quantum critical point. The
amplitude and width of this dip depend on structural parameters
that should be tunable in experiments to enhance the feature. This
opens interesting possibilities for controlled experimental study
of a quantum phase transition through conductance measurements.

\section{Conclusions}

We have analyzed the electronic transmission in two different
regimes of a double quantum-dot system. Utilizing an Anderson
Hamiltonian and its exact solution using numerical renormalization
group methods, one can determine the energy-dependent transmission
function for the structure. As different geometries are explored,
one can access an unusual Kondo regime with split peaks in the
spectral function, as well as a Kondo system in a pseudogapped
environment, allowing exploration of an interesting quantum
critical transition. The transmission function exhibits clear
signatures of these different Kondo regimes, opening the
possibility of extensive experimental studies of their properties
and response to different perturbations.


\end{document}